\documentclass[twoside,11pt]{8ssmmp} 

\usepackage{amsmath,amsfonts}

\usepackage{fancyhdr}    
\usepackage{graphicx}
\usepackage{rotating}

\newtheorem{definition}{Definition}

\newcommand{\diag}{\mathop{\mathrm{diag}}}

\begin{document}
 \baselineskip=11pt

\title{Dynamical Emergence of FRW Cosmological Models}
\author{\bf{Marek Szyd{\l}owski}\hspace{.25mm}\thanks{\,e-mail address: marek.szydlowski@uj.edu.pl}
\\ \normalsize{Astronomical Observatory, Jagiellonian University},
\\ \normalsize{Orla 171, 30-244 Krak\'{o}w, Poland} \\
\normalsize{Mark Kac Complex Systems Research Centre, Jagiellonian University}, 
\\ \normalsize{Reymonta 4, 30-059 Krak\'{o}w, Poland} \vspace{2mm} \\
\bf{Pawe{\l} Tambor}\hspace{.25mm}\thanks{\,e-mail address: pawel.tambor@gmail.com}
\\ \normalsize{Copernicus Center for Interdisciplinary Studies}, 
\\ \normalsize{Gronostajowa 3, 30-387 Krak\'{o}w, Poland} }

\date{}

\maketitle

\begin{abstract}

Recent astronomical observations strongly indicate that the current Universe is undergoing an accelerated phase of expansion. The discovery of this fact was unexpected and resulted in the comeback of cosmological constant. The conception of standard cosmological model has its roots in this context. The paper relates to the methodological status of effective theories in the context of cosmological investigations. We argue that the standard cosmological model (LCDM model) as well as the CDM have a status of effective theories only, similarly to the standard model of particle physics. The LCDM model is studied from the point of view of the methodological debate on reductionism and epistemological emergence in the science. It is shown in the paper that bifurcation as well as structural instability notion can be useful in the detection of emergence the LCDM model from the CDM model. We demonstrate that the structural stability of the LCDM model can explain the flexibility of the model to accommodation of the observational data. We show that LCDM model can be derived from CDM as the bifurcation. It is an example of acausal derivation of Lambda term. The case study of emergence of LCDM model suggests that it can be understood in terms of bifurcation and structural stability issue. The reduction from the upper models represented in terms of dynamical system to low-level ones can be realized in any case by application of a mathematical limit (boundary crossing) with respect to the model parameter. It is a simple consequence of mathematical theorem about smooth dependence solutions with respect to time, initial condition and the parameters.  

\end{abstract}

\clearpage 

\section{Introduction}
Recent astronomical observations of distant  type SNIa supernovae strongly indicate that the current Universe is undergoing an accelerated phase of expansion \cite{Perlmutter:1998np,Riess:2004nr}.
If the Universe's evolution is described by homogeneous and isotropic models filled with a perfect fluid, then the acceleration should be driven by a perfect fluid violating the strong energy condition. If different candidates for a fluid termed dark energy are suggested, the simple candidate for the dark energy in the form of positive cosmological constant seems to be the best one \cite{Kurek:2007tb}. While the Lambda CDM model is a suitable to phenomenologically describe the acceleration phase of the expansion of the Universe there is a serious problem with the interpretation of the Lambda term as a quantum vacuum energy due to the fine tuning problem. Our studies show that when the LCDM has the status of an effective theory, which offers description of the observational facts rather than their explanation, it introduces principally a new theoretical element which plays the role of an effective parameter changing the dynamics dramatically.

Our main result is that structural stability notion taken from the dynamical system theory may be useful in our understanding of the emergence of CDM to LCDM model as well as in understanding the reduction of LCDM to CDM one. We argue that the concepts of structural stability might be a suitable setup for the methodology of cosmology discussion and that the notion of bifurcation successfully replaces the notion of emergence. 

The LCDM model should in our opinion be treated as an effective theory for the following reasons. Firstly, theory of gravity which describes the gravitational sector of cosmology is very complicated, but if we postulate some simplified assumption like symmetry assumption idealization, then we obtain the simplest model which can be represented in the form of the dynamical system. In the cosmology, assumption of homogeneity and isotropy of space like sections of constant cosmic time ($t = const$) seems to be justified by the distribution of large scale structure of astronomical objects (cosmological principle). If we assume that universe is homogeneus and isotropic the evolution of it is determined by a single function of $a(t)$ called the scale factor. If we postulate that source of gravity is in the form of perfect fluid with energy density $\varrho (t)$ and pressure $p(t)$, then Einstein field equation reduces to the ordinary system of differential equation determining a single function $a(t)$. These equations called Friedmann equations FRW model describe the evolution of the Universe at the large scale.  Note that this is impossible basing on general relativity without the symmetry assumption where there is universal time conception. Finally, the general acceptance of the LCDM model as working is a good strategy \cite{Peebles:2007qe} but one may also seek alternative physics (pragmatism). 

\section{Structural stability issues}

If the space satisfies cosmological principle then Einstein field equations can be reduced to the system of ordinary differential equations, i.e. dynamical systems. Hence the dynamical system methods to the cosmology could be applied to cosmology in a natural way. The application of these methods allows us to reveal some stability properties of particular model visualized in geometrical way as the trajectories in the phase space. Therefore one can see how large the class of solutions leading to the desired property in tools of the attractors and the inset of limit set is (an attractor is a limit set with an open inset -- all the initial conditions that end up in the some equilibrium state). The attractors are the most prominent experimentally. It is because the probability of an initial state of the experiment to evolve asymptotically to the limit set is proportional to the volume of inset.

The idea, now called structural stability, emerged early in the history of dynamics investigation in 1930s the writings of Andronov, Leontovich and Pontryagin in Russia (1934) \cite{Andronov:1937} (the authors do not use the name \textit{structural stability}, but rather the name ''roughly systems''). This idea is based on an observation of an actual state of the system which can never be specified exactly and application of the dynamical systems might be useful anyway if it can describe the features of the phase portrait that persist when the state of the system is allowed to move around.

Among all dynamicists there are shared prejudices that: (1) there is a class of phase portraits that are far simpler than arbitrary ones. It explains why a considerable portion of the mathematical physics has been dominated by the search for the generic properties. The exceptional case should not arise very often in application and they de facto interrupt discussion (classification); (2) The physically realistic models of the world should possess some kind of the structural stability because having so many dramatically different models all agreeing with observation would be fatal for the empirical method of science (see also \cite{Thom:1977,Szydlowski:1984ss}.

These prejudices can in the Holton's terminology be treated as thematic principles \cite{Losee:1993,Holton:1981}.
In the cosmology a property (for example acceleration) is believed to be ''physically realistic'' if it can be attributed by the generic subsets of the models within a space of all admissible solutions or if it possesses a certain stability, i.e. if it is shared by a ''epsilon perturbed model''. 

The dynamical system is called structurally stable if all $\delta$-perturbation of it (sufficiently small) have the epsilon equivalent phase portrait. Therefore for the conception of structural stability we considered a $\delta$-perturbation of vector field determined by the right-hand sides of the system which is small (measured by delta). We also need a conception of the epsilon equivalence. It takes the form of topological equivalence--a homeomorphism of the state space preserving the arrow of time on each trajectory. In the definition, the structural stability considers only the deformation of ''rubber sheet'' type stretches or slides the phase space a small amount measured by epsilon. The main advantage of the structural stability is that it is the characterization of global dynamics itself.

Recently properties of structural stability of cosmological models were investigated by S. Kokarev \cite{Kokarev:2008ba}. In the introduction to the paper, the author claims that the history of cosmology shows that corrections of cosmological models are realized mainly by the sequence of their, in a sense ''small'', modifications and some of them may ''survive'' after small changes, while others may disappear. In the former case the property is referred to as ''rough'' or structurally stable, in the later one -- ''thin'' or structurally unstable. The author studies how some model properties, like singularities for example, will be present in the model if we ''perturb'' the model (e.g. generalize the Lagrangian of General Relativity). In our approach the property of structural stability is the property of the model itself. Also the type of perturbations is not specified (epsilon perturbation idea). Therefore, if we prove the structural instability of CDM model, the result will not depend on the choice of the type of perturbation. Then the property of structural stability becomes its constitutive property at the very beginning without restriction to the class of perturbation induced by considering new theories with generalized Lagrangian.

The idea of structural stability attempts to define the stability notion of differential deterministic models of the physical processes. In the case of planar dynamical systems (as in the case of models under consideration) there Peixoto theorem is true (Peixoto 1982) \cite{Peixoto:1962} which states that structurally stable dynamical systems form open and dense subsets in the space of all dynamical systems defined on the compact manifold. This theorem is a basic characterization of structurally stable dynamical systems in the plane which offers the possibility of the exact definition of generic (typical) and nongeneric (exceptional) cases (properties) by means of the notion of structural stability. Unfortunately there is no counterpart of this theorem in a more dimensional case when structurally unstable systems can also form open and dense subsets. For our aims, it is important that Peixoto theorem can give the characterization of generic cosmological models in terms of potential  function $V$ of the scale factor $a$ which determine the motion of the system of Newtonian type: $\ddot{a} = - \frac{\partial V}{\partial a}$.

Therefore we can treat FRW equation with various forms of dark energy as the two-dynamical system which looks like Newtonian type where the role of coordinate variable is played by the cosmological radius (or redshift $z$: $1+z = \frac{a_0}{a} \equiv x^{-1}$). We can construct an effective potential (the second order acceleration equation has exactly the Newtonian form) where the role of a coordinate variable is played by the cosmological radius.

Using the term of the structural stability first introduced by Andronov, Leontovich and Pontryagin in 1930s, one can classify different models of cosmic acceleration. It will be demonstrated that models with the accelerating phase which follow the deceleration are natural and typical from the point of view of the dynamical systems theory combined with the notion of structural stability in contrast to the models with bounces.

Let us introduce the following definition:

\begin{definition}
If the set of all vector fields $ \textbf{f} \in C^r (\mathcal{M})$ $(r \geq 1)$ having a certain property contains an open dense subset of $C^r (\mathcal{M})$, then the property is called generic.
\end{definition}

From the physical point of view it is interesting to know whether certain subset $v$ of $C^r (\mathcal{M})$ (representing a class of cosmological accelerating models in our case) contains a dense subset because it means that this property (acceleration) is typical in $V$.

It is not difficult to establish some simple relation between the geometry of potential function and the localization of critical points and its character for the case of dynamical systems of Newtonian type:
\begin{enumerate}
\item
The critical point of the system under consideration $\dot{x} = y$, $y = - \frac{\partial V}{\partial x}$ lies always on $x$-axis, i.e. they represent static universe $y_0 = 0$, $x=x_0$;
\item
The point $(x_0,0)$ is a critical point of the Newtonian system if it is a critical point of the potential function $V(x)$, i.e. $V(x) = E$ ($E$ is total energy of the system $E = \frac{y^2}{2} + V(x)$; $E = 0$ for the case flat models and $E = \frac{-k}{2}$ in general);
\item
If $(x_0,0)$ is a strict local maximum of $V(x)$, it is a saddle type critical point;
\item
If $(x_0,0)$ is a strict local minimum of the analytic function $V(x)$, it is the center;
\item
If $(x_0,0)$ is a horizontal inflection point of the $V(x)$, it is a cusp.
\end{enumerate}

Therefore the geometry of potential function determines the critical points as well as its stability. The integral of energy defines the algebraic curves in the phase space $(x,y)$ which are representing the evolution of the system with time. In any case the eigenvalues of the linearization matrix satisfy the characteristic equation $\lambda^2 + \frac{\partial^2 V}{\partial x^2}|_{x = x_0} = 0$.

\section{Cosmological models as dynamical systems}

Cosmology is based on the Einstein field equation which represents a very complicated system of partial nonlinear differential equations. Fortunately, the majority of main class of cosmological models from the point of view of observational data, belong to the class of the spatially homogeneous ones, for which the absolute cosmological time makes sense. As a consequence, the evolution of such models can be reduced to the systems of ordinary differential equations. Hence the methods of dynamical system theory or qualitative theory of differential equation can be naturally applied to cosmology. Among these classes of models especially interesting are the cosmological models with maximally symmetric space sections, i.e. homogeneous and isotropic. They are called FRW models (Friedmann-Robertson-Walker) if source of the gravity is a perfect fluid described in terms of energy density $\varrho$ and pressure $p$, both are the functions of cosmological time $t$. The FRW dynamics is described by two basic equations:
\begin{align}
\ddot{a} &= - \frac{1}{6} (\varrho + 3p)a = - \frac{\partial V}{\partial a} \\
\dot{\varrho} &= - 3 H (\varrho + p); ,
\end{align}
where the potential $V = - \frac{1}{6} \varrho a^2$, $a$ is the scale factor and $H = d\ln{a}/dt$ is Hubble's function, an overdot means the differentiation with respect to the cosmological time $t$.

The first equation is a consequence of the Einstein equations for the component (1,1), (2,2), (3,3) and the energy momentum tensor $T^{\mu}_{\nu} = \diag |-\varrho, p, p, p|$. This equation is called the Raychaudhuri or acceleration equation. The second equation represents the conservation condition $T^{\mu}_{\nu;\mu} = 0$.
It is very strange and unreasonable that such two simple equations satisfactorily describe the Universe's evolution at the large scales. Of course there is a more general class of cosmological models called the Bianchi models which have only the symmetry of homogeneity, but they fail to describe the current Universe which is isotropic as indicated measurement of the cosmic microwave background (CMB) radiation.

The system of equations (1) and (2) admits the first integral called the Friedmann equation
\begin{equation}
\varrho - 3H^2 = 3 \frac{k}{a^2},
\end{equation}
where $k$ is curvature constant $(0, \pm 1)$ and $\varrho$ plays the role of effective energy density.

If we consider the Lambda CDM model, then
\begin{equation}
\varrho_{\text{eff}} = \varrho_{\text{m},0} a^{-3} + \Lambda,
\end{equation}
i.e. energy density is the sum of dust matter (cold) and dark energy.
Therefore the potential function for the flat FRW model assumes the following form:
\begin{equation}
V = - \frac{\varrho_{\text{eff}} a^2}{6} = (-) \{\varrho_{\text{m},0} a^{-1} + \Lambda a^2\}.
\end{equation}

Formally the curvature effects as well as the cosmological constant term can be incorporated into the effective energy density ($\varrho_k = - \frac{k}{a^2}$; $\varrho_{\Lambda} = \Lambda$; $p_{\Lambda} = - \Lambda$, $p_{k} = - \frac{1}{3}\varrho_{k}$).

The form of equation (1) suggests the possible interpretation for the evolutional paths of cosmological models as a motion of a fictitious particle of unit mass in a one-dimensional potential parameterized by the scale factor. Following this interpretation the Universe is accelerating in the domain of configuration space $\{ a \colon a \geq 0 \}$ in which the potential is a decreasing function of the scale factor. In the opposite case if potential is a growing function of $a$ the Universe is decelerating. The limit case of zero acceleration corresponds to an extremum of the potential function.

\section{Emergence of the LCDM model from the CDM model in the framework of structural stability}

It should be mentioned that the notion of emergence applied to physics shows quite different results than in philosophy. We suggest that while having philosophical flavor, the emergence notion should be treated in philosophy of science with great caution. It appears to be used in physics rather in a colloquial and informal way and emergence in the context of the cosmology, as we show in this paper, does not mean irreducibility or unpredictability.  

The dynamical system investigation of the solutions of differential equations shifts its key point from founding and analyzing of individual solutions to investigating the space of all solutions for all admissible initial conditions, in the geometrical language of the phase space. Certain property (such as acceleration, singularities, etc.) is believed to be realistic if it can be attributed to a large subset of models within the space of all solutions \cite{Szydlowski:1984ss}. The evolutional scenarios are represented by the phase curves or by critical points, limit circles or other limit sets. We say that two dynamical systems (or equivalently vector fields), say $f(x)$ and $g(x)$) are equivalent, if there is an orientation preserving homeomorphism sending integral curves of $f$ into those of $g$. Of course this equivalence relation divided space of all dynamical systems on the plane on disjoint class of abstraction. Let phase space $E = R^n$, then $\epsilon$ - perturbation $f$ is the function $g \in C^1 (\mathcal{M})$ satisfying $\|f-g\|_1 < \epsilon$; where $\mathcal{M}$ is open subset of $R^n$ and $\|...\|_1$ is $C^1$ norm form the Banach space. In the introduced language it is natural to formulate an idea of structural stability. The intuition is very simple, namely $f$ is structurally stable vector field if for any vector field $f$ and $g$ are topologically equivalent. Then one can define the property of structural stability of the system.

\begin{definition} \label{def:2}
A vector field $f \in C^1 (\mathcal{M})$ is called structurally stable if there is an $\epsilon > 0$ such that for all $g \in C^1 (\mathcal{M})$ with $\|f-g\|_1 < \epsilon$, $f$ and $g$ are topologically equivalent on open subset $R^n$ called $\mathcal{M}$.
\end{definition}

The 2-dimensional case is distinguished by the fact that the Peixoto theorem (1962) gave a complete characterization of structurally stable systems on any compact, two dimensional space asserts that they are generic, i.e. forms open and dense subsets in the space of all dynamical system on the plane \cite{Peixoto:1962}.

While there is no counterpart to the Peixoto theorem in higher dimension, it can be easy used to test whether such dynamical systems or cosmological origin has a structurally stable global phase portrait. In particular, a vector field on the Poincar{\'e} sphere will be structurally unstable if there are non-hyperbolic critical points at infinity on the equator of the Poincar{\'e} sphere. In the opposite case if additionally the number of critical points and limit cycles is finite, $f$ is structurally stable on $S^2$.

In the next section we prove that the CDM model is structurally unstable (therefore exceptional in the space of all dynamical systems on the plane) and its transition (which we called emergence) to the Lambda CDM model means such perturbation of the CDM system that new perturbed system is structurally stable (therefore generic). 

\begin{figure}
\begin{center}
\includegraphics[width=0.3\textwidth]{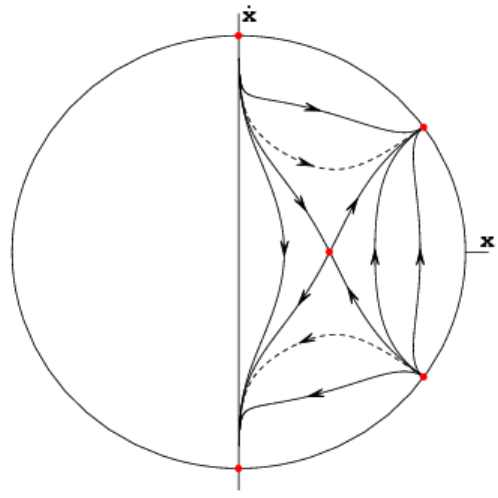}
\includegraphics[width=0.3\textwidth]{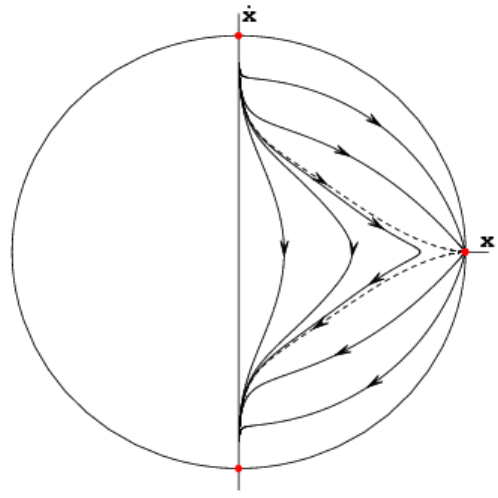}
\includegraphics[width=0.3\textwidth]{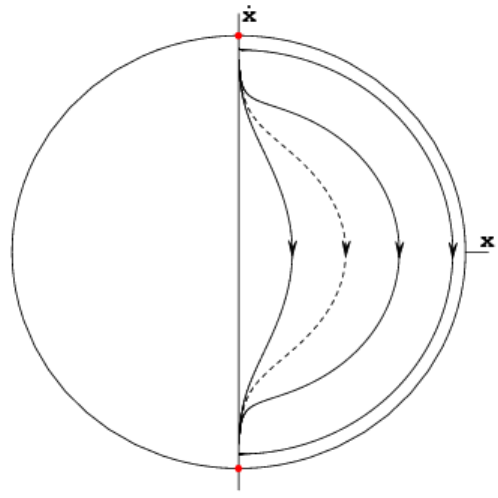}
\caption{The phase portraits for different perturbations of the CDM model -- from the left: 1) the LCDM model with the positive cosmological constant, 2) the LCDM model with the negative
cosmological constant, 3) the CDM model with the vanishing cosmological constant. Note that only the systems with the cosmological constant are structurally stable, while the CDM model is unstable because of the presence of degenerated critical points at the circle at infinity (case 3). The right figure represents the Einstein static universe $(x,\dot{x})=(\infty,0)$. The critical point $(0,\infty)$ represents the big-bang singularity (an unstable node).} \label{fig:3}
\end{center}
\end{figure}

\section{Emergence of new properties of the model through a bifurcation}

In our discussion it would be useful to consider a common approach to reduction in physics, so called \textit{deductive criterion of reducibility} of Nagel \cite{Nagel:1961en}. In this concept reduction is a relation of derivation between upper-level and base-level theories. This is exactly our point of view from the case study of cosmology as based on the dynamical systems theory that the dynamical model from upper-level can be smoothly reduced to the base-level one through the boundary crossing\footnote{Which is granted by theorems from differential equations theory about smooth dependence solutions on the time, initial conditions and the parameters.}. It is the example of a case where the emergence and reduction coexist.  

Let us consider two models which must be connected using the cosmological parameter. This parameter plays the role of control parameter in the model and we assume that it assumes zero (vanishes) in the basal model. We are looking for weakly emergent properties of the model which can be derived (via bifurcation) from the complete knowledge of the basal model information. In order to do so we use bifurcation theory, from which information about new unveiling properties of the system can be predicted, at least in principle as we change the control parameter. Then in principle we can derive the system behavior because we can perform bifurcation analysis answering the question how the structure of the phase space qualitatively changes as parameter $\Lambda$ is moved. Due to bifurcation we can not predict its future behavior with complete certainty (because there are to possible predictions concerning the value of cosmological constant; $\Lambda > 0$ or $\Lambda < 0$). Such a point of view seems to be very close to traditional conceptions of emergence (Broad, Popper, Nagel) that focus on unpredictability properties of upper-level model even given complete basal information. In the case of LCDM model novelty element is related with $\Lambda$-term, which emerges as a essential model parameter due to astronomical observation of type Ia distant supernovae (let us call it the empirical emergence). 

Let us illustrate our point of view in the a very simple, yet instructive, example. The dynamics of the flat cosmological models with R-W symmetry of space-like section, cosmological constant and without the matter (only for simplicity of presentation) is governed by a very simple equation (one-dimensional system)
\begin{equation}
\dot{x} = - x^2 + \frac{\Lambda}{3},
\end{equation}
where $x=H$ is Hubble parameter which measures the average rate of expansion of the Universe; $\Lambda$ is here the cosmological constant parameter; $\dot{}$ denotes differentiation with respect to the cosmological time.

Obviously, the above system can be simply integrated in quadratures. Calculation gives
\begin{equation}
x(t) = \sqrt{\frac{\Lambda}{3}} \coth \sqrt{\frac{\Lambda}{3}}(t+C)
\end{equation}
and $x(t=-C)= \infty$,
where $C$ is integration constant.
Equation (6) can be also integrated for the special case of $\Lambda=0$
\begin{equation}
x(t) = \frac{1}{t-t_0}.
\end{equation}

Naive looking at the formula (7) gives rise to statement that there is no transition from the solution (7) to (8) as $\Lambda \rightarrow 0$. It is not a true point of view. The limit solution as $\Lambda \rightarrow 0$ can be achieved in the following way. Let us consider $\Lambda$ in the formula (7) as small parameter and both functions $\cosh(...)$ i $\sinh(...)$ can be expanded in Taylor series with respect to $\sqrt{\frac{\Lambda}{3}}t$ term. Conserving linear parts of both functions, we obtain:
\begin{equation}
x(t) = \sqrt{\frac{\Lambda}{3}} \left[\frac{1+\sqrt{\frac{\Lambda}{3}}(t+C)+1-\sqrt{\frac{\Lambda}{3}}(t+C)}{1+\sqrt{\frac{\Lambda}{3}}(t+C)-1+\sqrt{\frac{\Lambda}{3}}(t+C)} \right]= \frac{1}{t+C}.
\end{equation}
Now, if we put $C=-t_0$, then the solution for CDM is reproduced. 
 
By this example one can observe how some small changes of the right hand side of the system dramatically change its solution. As a result in this system and solution emerge new asymptotic states representing de Sitter model.

\begin{figure}
\begin{center}
\includegraphics[]{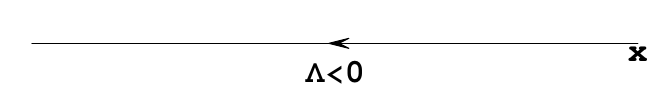}
\includegraphics[]{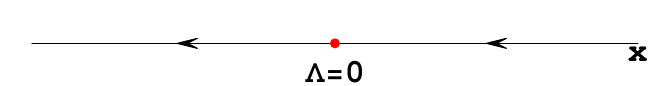}
\includegraphics[]{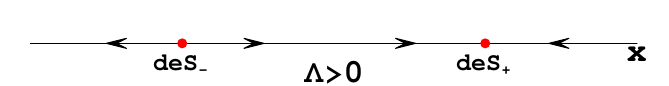}
\caption{The phase portraits for flat FRW model without matter and with the cosmological constant of different signs. For $\Lambda < 0$ there is no critical point. For $\Lambda = 0$ there a single degenerated critical point at the origin. For $\Lambda > 0$ there are two critical points, unstable deS${_{-}}$ and stable deS${_{-}}$.} \label{fig:7}
\end{center}
\end{figure}

The bifurcation theory serves to clarify the emergence of new properties (sometimes unexpected) of the system without solving this equation. Let us consider the system in the framework of bifurcation theory. For $\Lambda>0$ there are two critical points $\dot{x} = 0$ at $x \pm \sqrt{\frac{\Lambda}{3}}$. From the physical point of view they are representing de Sitter model (expanding and contracting). Derivative $f(x)$ ($\dot{x} = f(x)$), $Df(x, \mu) = -2x$ and $Df (\pm \sqrt{\frac{\Lambda}{3}}, \Lambda) = \mp 2 \sqrt{\frac{\Lambda}{3}}$l and we can see that the critical point at $x = \sqrt{\frac{\Lambda}{3}}$ is stable while the critical point $x = - \sqrt{\frac{\Lambda}{3}}$ is unstable. For $\Lambda = 0$, there is only one critical point at $x=0$ and it is a nonhypebolic critical point since $Df(0,0) = 0$; the vector field $f(x) = -x^2$ is structurally unstable; $\Lambda = 0$ is a bifurcation value. For $\Lambda <0$ there are no critical points. The phase portraits for this differential equation are shown in Fig.~\ref{fig:7}.

\begin{figure}
\begin{center}
\includegraphics[width=0.7\textwidth]{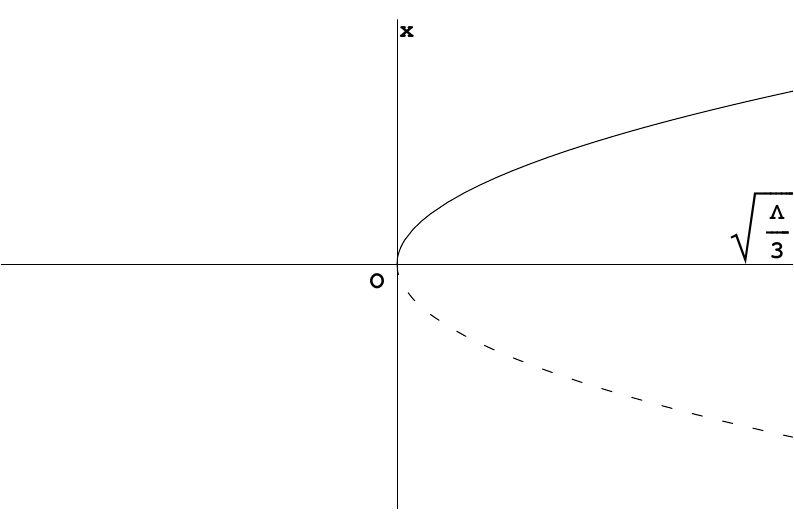}
\caption{The bifurcation diagram of the flat LDM model (LCDM model without matter). It is the fork bifurcation of saddle-node type. The Einstein universe ($x=0$) bifurcates to the expanding (upper stable branch) and contracting (lower unstable branch) de Sitter models.} \label{fig:8}
\end{center}
\end{figure}

In this case we have $W^{s} (\sqrt{\frac{\Lambda}{3}}) = (-\sqrt{\frac{\Lambda}{3}}, \infty)$ and $W^{u} (-\sqrt{\frac{\Lambda}{3}}) = (- \infty, \sqrt{\frac{\Lambda}{3}})$ as, respectively, a stable and unstable manifold . And for $\Lambda = 0$ the one--dimensional center manifold is given by $W^{c} (0) = (- \infty, \infty)$. All of the pertinent information concerning the bifurcation that takes place in this system at $\Lambda = 0$ is captured in the bifurcation  diagram shown in Fig.~\ref{fig:8}. The curve $\frac{\Lambda}{3} - x^2 = 0$ determines the position of the critical points of the system, a solid curve is used to indicate a family of stable critical points while a dashed curve is used to indicate a family of unstable critical points. This type of bifurcation is called a saddle-mode bifurcation.

The system under consideration constitutes only an example of dynamical system analysis of the system cosmological origin but there are many other systems with some parameter which shows hidden and unexpected properties as parameter varies. Let us remind some of them. In the problem of the motion star around the elliptic galaxy appears Henon, Heiles \cite{Henon:1964} Hamiltonian system. This system possesses the energy first integral $E$ and if $E > E_{crit}$ then transition to the chaotic behavior appears. Another example of bifurcation and emergence of the cyclic behavior in the system of the limit cycle type is offered by the famous van der Pol equation $\ddot{x} + \mu (x^2 - 1) \dot{x} + x = 0$. For $\mu = 0$ the system is of harmonic oscillator type and for $\mu >0$, van der Pol's equation has a unique limit cycle and is stable \cite{Perko:2001}. The limit cycle is representing a closed trajectory in the phase space which attracts all trajectories from neighborhood.

In this case $\epsilon = 0$ is a bifurcation value parameter and limit cycle behavior is an upper-level emergent property. For the interesting discussion on emergence, basal and upper-level models and reducibility see \cite{Wayne:2008}.

Also interesting experiences of emergence new type of dynamical behavior are provided by Hopf bifurcation phenomena \cite[s. 341]{Perko:2001}. This bifurcation can occur in the system with parameter $\dot{x} = f(x, \mu)$ at a nonhyperbolic equilibrium point $x_0$ when the matrix $Df(x_0, \mu_{0})$ has a simple pair of pure imaginary eigenvalue and no other eigenvalues with zero real point. In the generic case Hopf bifurcation occurs where a periodic orbit is created as the stability of equilibrium point $x_{\mu}$ changes. This type of behavior plays an important role in the description route to turbulence scenario. It would be worth mentioning the important role of Hopf bifurcation in Rulle-Takens scenario of route to deterministic chaos. The concept of turbulence war originally introduced by Landau in 1944 and later revised by Ruelle and Takens in 1941 \cite{Cvitanovic:1989}.
According to Landau, turbulence is reached at the end of an indefinite superposition of oscillatory bifurcations, each bringing its unveiling phase into the dynamics of the system. In the Ruelle-Takens scenario infinite number of periodic behavior is not required when nonlinearities act. They argue that turbulence should be treated as a stochastic regime of deterministic chaos at which long term unpredictability occurs due to property of sensitive dependence on initial condition. This stage is reached only after a finite and small number of bifurcations. For some recent philosophical discussion on the significance of chaos see \cite{Werndl:2008}.

\section{Conclusion}

The main aim of this paper was to show the effectiveness of using the framework of dynamical system theories (especially the notion of structural stability and tools of bifurcation analysis) in the study on emergence and reducibility of two cosmological models Cold Dark Matter model and Lambda Cold Dark Matter model. 

We have also demonstrated reducibility of solutions of LCDM upper state after taking the limit $\Lambda \rightarrow 0$. If we include the theorem about smooth dependence of solutions on initial condition and parameters, one can obtain CDM model as the boundary crossing. The analogical problem appears in the Wayne analysis of limit cycle behavior emergence in nonlinear system. In this case $\epsilon =0$ is bifurcation value of parameter. In our interpretation space of state of the system can be parametrized by epsilon parameter which measures the strength of the nonlinear term. As a consequence bifurcation analysis reveals a new type of dynamical behavior for any value of epsilon parameter. This upper state can be reduced to the basal state because it is guaranteed by theorems mentioned above, which we applied in the context of LCDM model. We can find a strict analogy to the system under consideration of cosmological origin and Wayne analysis of emergence and singular limits.
Following the common approach to reductionism in physics, so called deductive criterion of reducibility by Nagel (1961), the reduction is a derivational relation between upper-level and base-level theories.
The structural instability of the models teaches us that one should not distinguish the derivational relation either on the level of basic equation and on the level of solutions. 
In the mathematical modeling of physical processes, we always try to convey the features of typical, garden-variety, dynamical systems. In mathematics the exceptional cases are more complicated and numerous, and they interrupt the physical discussion. Moreover, dynamicists share an opinion that such exceptional systems do not arise very often because they are not typical. The history of mathematical dynamics presents the search for generic properties. We would like to distinguish a class of phase portraits that are far simpler than the arbitrary ones. This program was achieved for dynamical systems on the plane by Peixoto due to the conception of structural stability introduced in 1934 by Andronov and Leontovich. The criteria for structural stability rely upon two supplementary notions: a perturbation of the phase portraits (or vector field) and the topological equivalence (homeomorphism of the state phase). A phase portrait has the property of structural stability if all sufficiently small perturbations of it have equivalent phase portraits. For example, one considers a center type of critical points, then the addition of perturbation pointing outward results in a point repellor which is not topologically equivalent to the center. This is a primary example of structurally unstable system. In the opposite case saddle type of critical point is structurally stable and the phase portrait does not change under small perturbation.

In this paper we define the class of FRW cosmological models filled by dark energy as two-dimensional dynamical systems of a Newtonian type. They are characterized through the single smooth effective potential function of the scale factor or redshift. Among these classes of models we distinguish typical (generic) and exceptional (nongeneric) cases with the help of structural stability notion and the Peixoto theorem. We find that the LCDM model is structurally stable as opposed to the CDM model. We demonstrate that this model represents a typical structurally stable perturbation of CDM one. Therefore, the transition from the CDM model of the Universe toward the LCDM one, which includes the effects of the cosmological constant, can be understood as an emergence of the model from the exceptional case to the generic one. This case represents a generic model in this sense that small changes of its right-hand sides do not change the global phase portraits. In terms of the potential, the second order differential equation, one can classify different models of cosmic acceleration. It is shown that models with the accelerating phase (following the deceleration) are natural and typical from the point of view of the dynamical systems theory combined with the notion of structural stability.

From the dynamical system theory we know that the solution of a dynamical system smoothly depends on the time, initial condition and the parameters. As a result we obtain that in any case as the model is represented by dynamical system there is a reduction from the upper-level model to the lower-level one by taking the mathematical limit with respect to the parameter. In the case of cosmology, the corresponding limit is obtained as the cosmological constant parameter goes to zero.

\subsection*{Acknowledgements}
The work was supported by the grant NCN DEC-2013/09/B/ST2/03455.


\begin{thebibliography}{B-B} 
\medskip
\begin{footnotesize} 

\bibitem{Perlmutter:1998np}
 {S.}~{Perlmutter}
   {et~al.} ({Supernova Cosmology Project}),
   {Astrophys. J.} \textbf{ {517}},
   {565} ({1999}), \textit{eprint}: {astro-ph/9812133}.

\bibitem{Riess:2004nr}
 {A.~G.}  {Riess} {et~al.}
  ({Supernova Search Team}),
   {Astrophys. J.} \textbf{607},
   {665} ({2004}), \textit{eprint}: {astro-ph/0402512}.

\bibitem{Lukash:2007ns}
 { {V.}~{Lukash}},
   {Nuovo Cim.} \textbf{ {122B}},
   {1411} ({2007}), \textit{eprint}: {0712.3356}.

\bibitem{Peebles:2007qe}
 { {P.~J.~E.}  {Peebles}},
   {Nuovo Cim.} \textbf{ {122B}},
   {1035} ({2007}), \textit{eprint}: {0712.2757}.

\bibitem{Andronov:1937}
 { {A.~A.}  {Andronov}}  {and}
   { {L.~S.}  {Pontryagin}},
   {Dokl. Acad. Sci. URSS} \textbf{ {14}},
   {247} ({1937}).

\bibitem{Thom:1977}
 {R.}~{Thom},
  \emph{ {Stabilit\'{e} structurelle et morphog\'{e}n\'{e}se}}
  ({Inter\'{e}dition}, {Paris},
   {1977}).

\bibitem{Szydlowski:1984ss}
 {M.}~{Szydlowski},
   { {M.}~{Heller}},  {and}
   { {Z.}~{Golda}},
   {Gen. Relat. Grav.} \textbf{ {16}},
   {877} ({1984}).

\bibitem{Losee:1993}
 {J.}~{Losee},
  \emph{ {A historical introduction to the philosophy of
  science}} ({Oxford University Press},
   {Oxford},  {1993}).

\bibitem{Holton:1981}
 {G.}~{Holton}, in
  \emph{{Scientific Explanation}}, edited by
  { {A.~F.}  {Heath}}
  ({Clarendon Press}, {Oxford},
   {1981}).

\bibitem{Kokarev:2008ba}
 {S.~S.}  {Kokarev}
  ({2008}), \textit{eprint}: {0810.5080}.

\bibitem{Peixoto:1962}
 {M.}~{Peixoto},
   {Topology} \textbf{ {1}},
   {747} ({1962}).

\bibitem{Peebles:1993}
 {P.~J.~E.}  {Peebles},
  \emph{ {Principles of Physical Cosmology}}
  ({Princeton University Press},
   {Princeton}, {1993}).

\bibitem{Nagel:1961en}
 {E.}~{Nagel},
  \emph{The Structure of science} ({Hackett
  Publishing Company, Incorporated}, {New York},
   {1961}).

\bibitem{Henon:1964}
 {M.}~{Henon}  {and}
   { {C.}~{Heiles}},
   {Astron. J.} \textbf{69},
   {73} ({1964}).

\bibitem{Perko:2001}
 {L.}~{Perko},
  \emph{Differential Equations and Dynamical Systems (Texts in
  Applied Mathematics, Vol 7)} ({Springer-Verlag New York},
   {New York},  {2001}).

\bibitem{Wayne:2008}
 {A.}~{Wayne} ({2008}),
  	extit{eprint}: {http://philsci-archive.pitt.edu/archive/00003933}.

\bibitem{Cvitanovic:1989}
 {P.}~{Cvitanovic}, in
  \emph{Universality in Chaos, 2nd ed}, edited by
  {P.}~{Cvitanovic}
  ({Taylor \& Francis},  {1989}).

\bibitem{Werndl:2008}
 {C.}~{Werndl}
  ({2008}),
  	\textit{eprint}: {http://philsci-archive.pitt.edu/archive/00003914}.

\bibitem{Kurek:2007tb}
 {A.}~{Kurek} {and}
   {M.}~{Szydlowski},
   {Astrophys. J.} \textbf{ {675}},
   {1} ({2008}), 	\textit{eprint}: {astro-ph/0702484}.

\bibitem{Bishop:2006}
 {R.~C.} {Bishop} {and}
   { {H.}~{Atmanspacher}},
   {Found. Phys.} \textbf{ {36 12}},
   {1757} ({2006}).

\end{footnotesize} 
\end{thebibliography}
\end{document}